\documentclass{ws-procs9x6}
    \usepackage{latexsym,bm,amsmath,amssymb,amsfonts}
    \usepackage{epsfig,graphics,graphicx}
    \usepackage{makeidx}
    \usepackage{citesort}

\newcommand{\rmd}{{\rm d}}   %ELS%

  \long\def\comment#1{ }
  \newcommand{\dif}{{\rm d}}
  \newcommand{\dY}{\dif Y}

  \newcommand{\del}{\partial}
  \newcommand{\lan}{\langle}
  \newcommand{\ran}{\rangle}
  \newcommand{\mcal}{\mathcal}
  \newcommand{\rme}{{\rm e}}

  \newcommand{\order}[1]{\mcal{O}{(#1)}}

  \newcommand{\beq}{\begin{eqnarray}}
  \newcommand{\eeq}{\end{eqnarray}}
  
 \def\simge{\mathrel{%
   \rlap{\raise 0.511ex \hbox{$>$}}{\lower 0.511ex \hbox{$\sim$}}}}
\def\simle{\mathrel{
   \rlap{\raise 0.511ex \hbox{$<$}}{\lower 0.511ex \hbox{$\sim$}}}}

\begin{document}

\title{In the Shadow of the Color Glass\footnote{\uppercase{T}his contribution
combines two talks presented by the author at \uppercase{DIS2006} in
the plenary session and, respectively, in the parallel session on
\uppercase{D}iffraction and \uppercase{V}ector \uppercase{M}esons.}}

\author{Edmond Iancu}

\address{Service de Physique Theorique, CEA Saclay, CEA/DSM/SPhT,\\
F-91191 Gif--sur--Yvette, France. E-mail: eiancu@cea.fr}

\maketitle

\abstracts{I give a brief overview of recent theoretical progress
within perturbative QCD concerning the high--energy dynamics in the
vicinity of the unitarity limit. Special attention is payed to the
most recent developments concerning the relation between
high--energy QCD and statistical physics, the role of the pomeron
loops, and the transition from geometric scaling to diffusive
scaling with increasing energy.}

\section{Motivation: The rise of the gluon distribution at HERA}

\begin{figure}
\begin{center}%\hspace*{-.3cm}
       \mbox{{\epsfig{figure=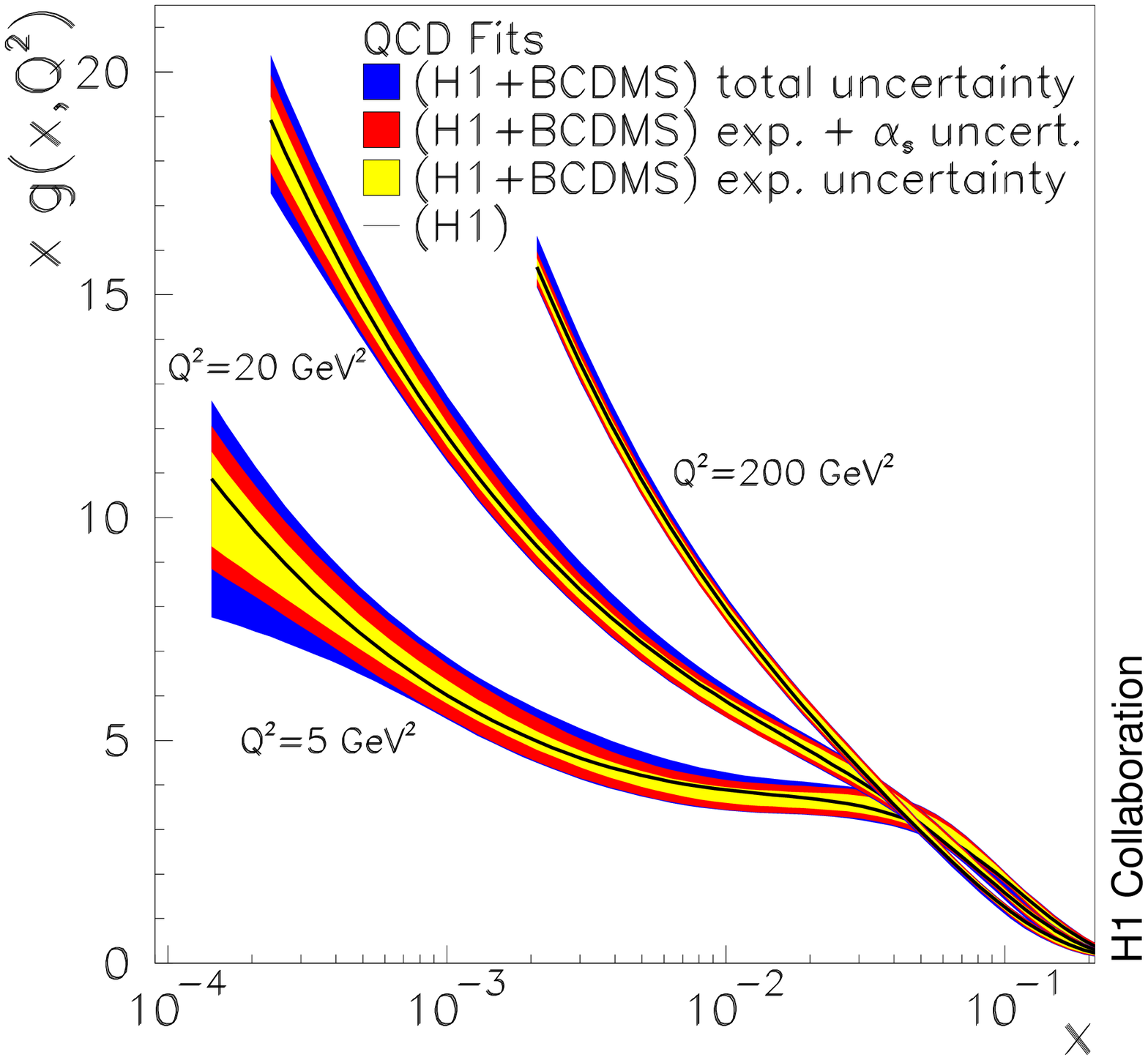,
        width=0.47\textwidth}}\hspace*{-.1cm}
             {\epsfig{file=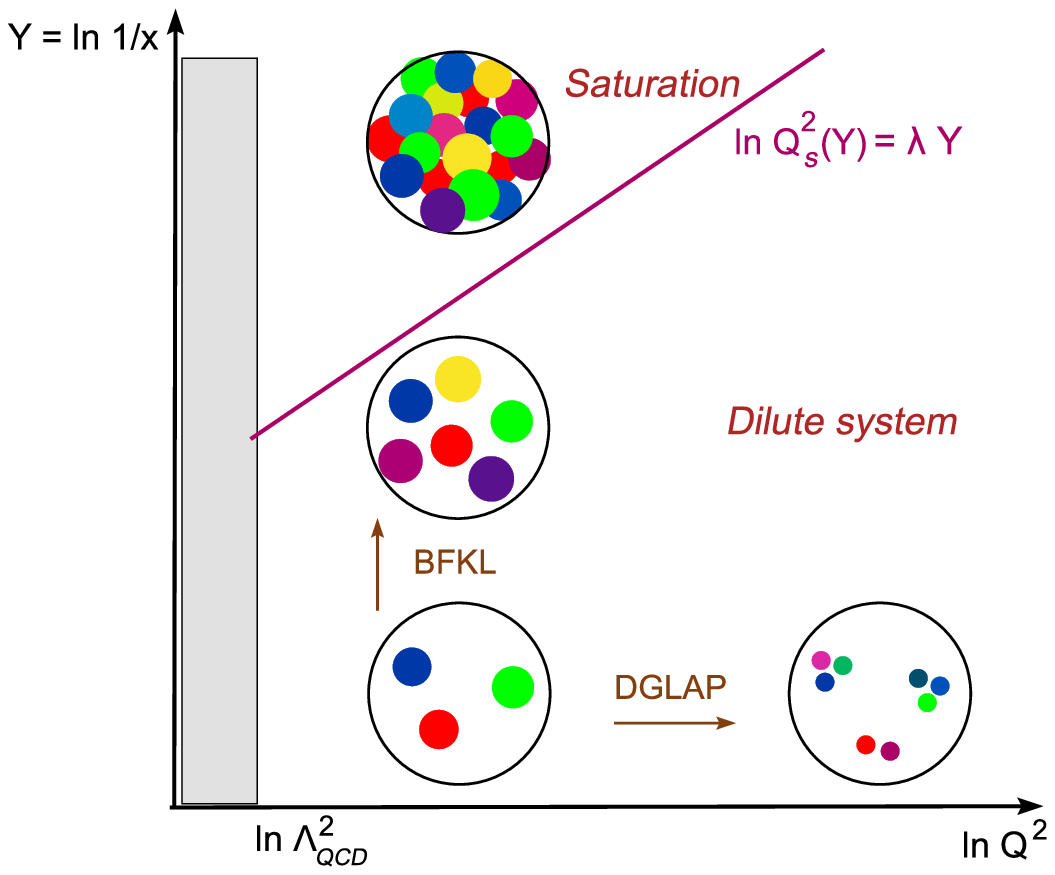,
        width=0.53\textwidth}}}
 %\centerline{\epsfig{figure=HERAgluon.eps,
%        width=0.65\textwidth}}
\caption{\sl Left: Gluon distribution extracted at HERA (here, data
from H1), as a function of $x$ in three bins of $Q^2$. Right: The
`phase--diagram' for QCD evolution suggested by the HERA data; each
colored blob represents a parton with transverse area $\Delta
x_\perp \sim 1/Q^2$ and longitudinal momentum $k_z=xP_z$.
\label{HERA-gluon}}
\end{center}\vspace{-.4cm}
\end{figure}

The essential observation at the basis of the recent theoretical
progress in the physics of hadronic interactions at high energy is
the fact that high--energy QCD is the realm of high parton (gluon)
densities and hence it can be studied from first principles, via
weak coupling techniques. Anticipated by theoretical developments
like the BFKL equation\cite{BFKL} and the GLR
mechanism\cite{GLR,MQ85} for gluon saturation, this observation has
found its first major experimental foundation in the HERA data for
electron--proton deep inelastic scattering (DIS) at small--$x$. As
visible, e.g., on the H1 data shown in Fig. \ref{HERA-gluon} (left
figure), the gluon distribution $xG(x,Q^2)$ rises very fast when
decreasing Bjorken--$x$ at fixed $Q^2$ (roughly, as a power of
$1/x$), and also when increasing $Q^2$ at a fixed value of $x$. The
physical interpretation of such results is most transparent in the
proton infinite momentum frame, where $xG(x,Q^2)$ is simply the
number of the gluons in the proton wavefunction which are localized
within an area $\Delta x_\perp \sim 1/Q^2$ in the transverse plane
and carry a fraction $x=k_z/P_z$ of the proton longitudinal
momentum.

Thus, without any theoretical prejudice, the HERA data suggest the
physical picture illustrated in the right hand side of Fig.
\ref{HERA-gluon}, which shows the distribution of partons in the
transverse plane as a function of the kinematical variables for
DIS in logarithmic units: $\ln Q^2$ and $Y\equiv \ln(1/x)$. The
number of partons increases both with increasing $Q^2$ and with
decreasing $x$, but whereas in the first case (increasing $Q^2$)
the transverse area $ \sim 1/Q^2$ occupied by every parton
decreases very fast and more than compensates for the increase in
their number --- so, the proton is driven towards a regime which
is more and more dilute ---, in the second case (decreasing $x$)
the partons produced by the evolution have roughly the same
transverse area, hence their density is necessarily increasing.

Accordingly, the DGLAP equation\cite{DGLAP} which describes the
evolution with increasing $Q^2$ is naturally {\em linear}, and also
{\em local} in $Q^2$. By contrast, the BFKL equation, which is the
linear equation originally proposed\cite{BFKL} to describe the
evolution with increasing energy, is {\em non--local} in transverse
space and should be merely regarded as a linear approximation to
more general evolution equations which are {\em non--linear}, i.e.,
which account for the interactions among the partons within the
wavefunction. The non--linear effects are expected to become
important in the region denoted as `saturation' in Fig.
\ref{HERA-gluon}, and also in the approach towards it when coming
from the dilute region at large $Q^2$.

Mainly because of its complexity, the high--energy evolution in QCD
is not as precisely known as the corresponding evolution with $Q^2$.
Still, the intense theoretical efforts over the last years led to
important conceptual clarifications and to new, more powerful,
formalisms --- among which, the effective theory for the Color Glass
Condensate (CGC)~\cite{MV,CGC,EdiCGC}
---, which encompass the non--linear dynamics in high--energy QCD
to lowest order in $\alpha_s$ and allow for a unified picture of
various high--energy phenomena ranging from DIS to heavy--ion, or
proton--proton, collisions, and to cosmic rays.

These developments may explain some remarkable phenomena observed in
the current experiments (like the `geometric scaling' in the HERA
data at small $x$\cite{geometric,MS06} and the particle production
at forward rapidities in deuteron--gold collisions at
RHIC\cite{Brahms-data}), and, moreover, they have potentially
interesting predictions for the physics at LHC. It is my purpose in
what follows to provide a brief, pedagogical, introduction to such
new ideas, with emphasis on the physical picture and its
consequences for deep inelastic scattering at high energy.

\section{DIS: Dipole factorization \& Saturation momentum}
\label{sec:dipole}

At small $x$, DIS is most conveniently computed by using the {\em
dipole factorization} (see, e.g., Refs.\cite{EdiCGC} for more
details and references). The small--$x$ quark to which couple the
virtual photon is typically a `sea' quark produced at the very end
of a gluon cascade. It is then convenient to disentangle the
electromagnetic process $\gamma^*q$, which involves this `last'
emitted quark, from the QCD evolution in the proton, which involves
mostly gluons. This can be done via a Lorentz boost to the `dipole
frame' in which the struck quark appears as an excitation of the
virtual photon, rather than of the proton. In this frame, the proton
still carries most of the total energy, while the virtual photon has
just enough energy to dissociate long before the scattering into a
`color dipole' (a $q\bar q$ pair in a color singlet state), which
then scatters off the gluon fields in the proton. This leads to the
following factorization:
  \beq\label{dipolefact}
  \sigma_{\gamma^*p}(x,Q^2)
 \, = \, \int_0^1 \rmd z \int \rmd^2 r\ \vert \Psi_{\gamma}(z,r;
 Q^2)\vert^2 \,\sigma_{\rm dipole}(x,r)
 \eeq
where $\vert \Psi_{\gamma}(z,r; Q^2)\vert^2$ is the probability for
the $\gamma^*\to q\bar q$ dissociation ($r$ is the dipole transverse
size and $z$ the longitudinal fraction of the quark), and
$\sigma_{\rm dipole}(x,r)$ is the total cross--section for
dipole--proton scattering and represents the hadronic part of DIS.
At high energy, the latter can be computed in the eikonal
approximation as
 \beq  \sigma_{\rm dipole}(x,r)\ = \ 2\int \rmd^2b\ \ T(r,b,Y)
 \,\eeq
where $T(r,b,Y)$ is the {\em forward scattering amplitude} for a
dipole with size $r$ and impact parameter $b$. This is the quantity
that we shall focus on. The unitarity of the $S$--matrix requires
$T\le 1$, with the upper limit $T=1$ corresponding to total
absorbtion, or `black disk limit'.

But the unitarity constraint can be easily violated by an incomplete
calculation, as we demonstrate now on the example of lowest--order
(LO) perturbation theory. To that order, $T(r,b,Y)$ involves the
exchange of two gluons between the dipole and the target. Each
exchanged gluon brings a contribution $gt^a \bm{r}\cdot\bm{E}_a$,
where $\bm{E}_a$ is the color electric field in the target. Thus,
$T\sim g^2 r^2 \langle \bm{E}_a\cdot\bm{E}_a\rangle_x$, where the
expectation value is recognized as the number of gluons per unit
transverse area:
  \beq\label{T1SCATT}
    T(x,r,b) \,\sim\,
  \alpha_s\,r^2\,\frac{xG(x,1/r^2)}{\pi
  R^2}\,\equiv\, \alpha_s\,n(x, Q^2\sim 1/r^2)\,.\eeq
In the last equality we have identified the {\em gluon occupation
number}\,: $n(x,Q^2)=$ [number of gluons $xG(x,Q^2)$] times [the
area $1/Q^2$ occupied by each gluon] divided by [the proton
transverse area $\pi R^2$].

Eq.~(\ref{T1SCATT}) applies so long as $T\ll 1$ and shows that weak
scattering (or `color transparency') corresponds to low gluon
occupancy $n\ll 1/\alpha_s$. But if naively extrapolated to very
small values of $x$, this formula leads to {\em unitarity
violations} : $T$ would eventually become larger than one ! Before
this happens, however, new physical phenomena are expected to come
into play and restore unitarity. As we shall see, these are {\em
non--linear} phenomena, and are of two types: \texttt{(i)} {\em
multiple scattering}, i.e., the exchange of more than two gluons
between the dipole and the target, and \texttt{(ii)} {\em gluon
saturation}, i.e., non--linear effects in the proton wavefunction
which tame the rise of the gluon distribution at small~$x$.

Eq.~(\ref{T1SCATT}) also provides a criterion for the onset of
unitarity corrections: These should become important when
$T(x,r)\sim 1$ or $n(x,Q^2)\sim 1/\alpha_s$. This condition can be
solved for the {\em saturation momentum}, which is the value of the
transverse momentum below which saturation effects are expected to
be important in the gluon distribution. One thus finds
 \beq\label{Qsat}
   Q^2_s(x)\, \simeq \,
{\alpha_s }\, \frac{x G(x,Q^2_s)}{\,\pi R^2}\, \,\,\sim\,\,
 \frac{1}{x^{\lambda}}\,,\eeq
which grows with the energy as a power of $1/x$, since so does the
gluon distribution before reaching saturation. In logarithmic units,
the {\em saturation line} $\ln Q^2_s(Y)=\lambda Y$ is therefore a
{\em straight} line, as illustrated in the right hand side of Fig.
\ref{HERA-gluon}. This is the borderline between the dilute regime
at high transverse momenta $k_\perp\gg Q_s(Y)$, where one expects
the standard perturbation theory to apply, and a high--density
region at low momenta $k_\perp\simle Q_s(Y)$, where physics is
non--linear. In fact, as we shall argue below, at high energy the
effects of saturation can extend up to very high values of
$k_\perp$, well above the saturation line.

\section{BFKL evolution: The blowing--up gluon distribution}
\label{sec:BFKL}

\begin{figure}
\begin{center}
\centerline{\epsfig{file=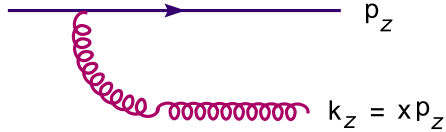,height=1.8cm}} \caption{\sl
The lowest order diagram for bremsstrahlung.\label{OneGluon}}
\end{center}\vspace*{-.8cm}
\end{figure}

\begin{figure}
\begin{center}
\centerline{\epsfig{file=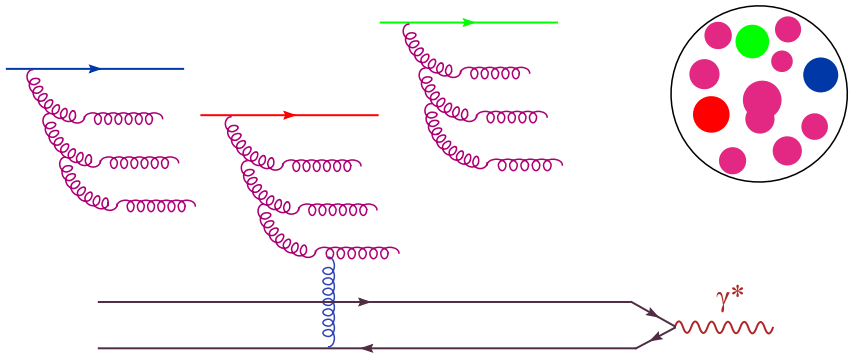,width=0.8\textwidth}}
\caption{\sl Gluon cascades produced by the high--energy (BFKL)
evolution of the proton wavefunction, as probed by a small color
dipole in DIS.\label{BFKLfig}}
\end{center}\vspace*{-.8cm}
\end{figure}

Within perturbative QCD, the emission of small--$x$ gluons is
amplified by the infrared sensitivity of the bremsstrahlung process,
whose iteration leads to the BFKL evolution (at least, for not too
high energies). Fig. \ref{OneGluon} shows the emission of a gluon
which carries a fraction $x=k_z/p_z$ of the longitudinal momentum of
its parent quark. When $x\ll 1$, the differential probability for
this emission can be estimated as
 \beq\label{brem} \rmd P_{\rm Brem}\,\simeq\,\frac{\alpha_s
 C_F}{2\pi^2}\,\frac{\rmd^2k_\perp}{k_\perp^2}\,\frac{\rmd
 x}{x}\,,\eeq
which is singular as $x\to 0$.
Introducing the rapidity $Y\equiv \ln(1/x)$, and hence $\rmd Y=\rmd
x/x$, Eq.~(\ref{brem}) shows that there is a probability of
$\order{\alpha_s}$ to emit one gluon per unit rapidity. The same
would hold for the emission of a soft photon from an electron in
QED. However, unlike the photon, the child gluon is itself charged
with `colour', so it can further emit an even softer gluon, with
longitudinal fraction $x_1=q_z/k_z\ll 1$. When the rapidity is
large, $\alpha_s Y\gg 1$, such successive emissions lead to the
formation of gluon cascades, in which the gluons are ordered in
rapidity and which dominate the small--$x$ part of the hadron
wavefunction (see Fig. \ref{BFKLfig}).

So long as the density is not too high, the gluons do not interact
with each other and the evolution remains {\em linear} : when
further increasing the rapidity in one more step ($Y\to Y+\dY$), the
gluons created in the previous steps {\em incoherently}  act as {\em
color sources} for the emission of a new gluon. This picture leads
to the following, schematic, evolution equation
 \beq\label{BFKL}
 \frac{\partial n}{\partial Y}\,\simeq\,\omega\alpha_s n \qquad
 \Longrightarrow \quad n(Y)\,\propto\, {\rm e}^{\omega\alpha_s Y}\,,
 \eeq
which predicts the exponential rise of $n$ with $Y$. This is an
oversimplified version of the BFKL (Balitsky-Fadin-Kuraev-Lipatov)
equation\cite{BFKL} which captures the main feature of this
evolution: the unstable growth of the gluon distribution. One knows
by now that this growth is considerably tempered by NLO
effects\cite{NLBFKL,DISNLO}, like the running of the QCD coupling or
the requirement of energy conservation, but the basic fact that the
gluon density increases exponentially with $Y$ is expected to remain
true (independently of the order in $\alpha_s$) so long as one
neglects the {\em non--linear} effects, or `gluon saturation', in
the evolution.

\section{Non--linear evolution: JIMWLK equation
and the CGC} % and the Color Glass Condensate}

\begin{figure}
 \begin{center}
 \centerline{\epsfig{file=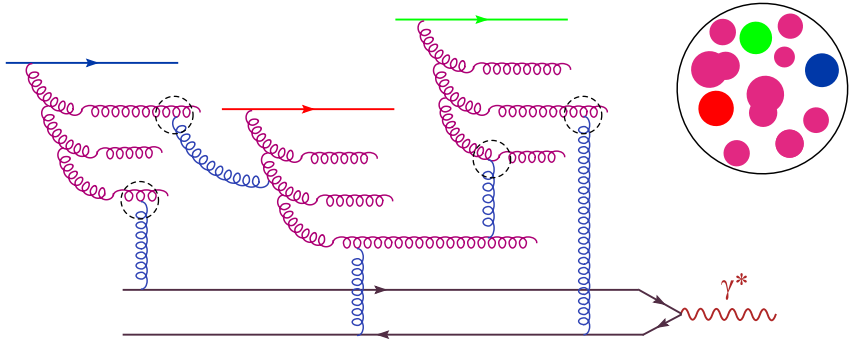,width=0.8\textwidth}}
   \caption{\sl  DIS in the presence of BFKL evolution, saturation
   and multiple scattering.\label{EvolREC}}
            \end{center}\vspace*{-.8cm}
\end{figure}

Non--linear effects appear because gluons carry colour charge,
so they can interact with each other % within the wavefunction
(even when separated in rapidity) by exchanging gluons in the
$t$--channel, as illustrated in Fig. \ref{EvolREC}. These
interactions are amplified by the gluon density and thus they should
become more and more important when increasing the energy.

Back in 1983, L. Gribov, Levin and Ryskin\cite{GLR} suggested that
gluon saturation should proceed via $2\to 1$ `gluon recombination',
which is a process of order $\alpha_s^2 n^2$ (cf. Fig.
\ref{EvolREC}). To take this into account, they proposed the
following, {\em non--linear}, generalization\footnote{This equation
has been later justified within perturbative QCD, by Mueller and
Qiu\cite{MQ85}, but only by performing some drastic approximations.}
of Eq.~(\ref{BFKL}) :
 \beq\label{GLR}
  \frac{\partial n}{\partial Y}\,\simeq\,\alpha_s n\,-\,
{\alpha_s^2}\, n^2 \, = \, 0 \quad {\rm when}\quad n\,
 \,\sim\,\frac{1}{\alpha_s}\,\gg\,1
 \eeq
which has a fixed point $n_{\rm sat}=1/\alpha_s$ at high energy, as
indicated above. That is, when $n$ is as high as $1/\alpha_s$, the
emission processes (responsible for the BFKL growth) are precisely
compensated by the recombination ones, and then the gluon occupation
factor saturates at a fixed value.

Twenty years later, we know that the actual mechanism for gluon
saturation in QCD is more subtle than just gluon recombination and
that its mathematical description is considerably more involved than
suggested by Eq.~(\ref{GLR}). This mechanism, as encoded in the
effective theory for the CGC and its central equation, the JIMWLK
equation (Jalilian-Marian, Iancu, McLerran, Weigert, Leonidov, and
Kovner)\cite{JKLW,CGC,W}, is the {\em saturation of the gluon
emission rate due to high density effects} : At high density, the
gluons are not independent color sources, rather they are strongly
correlated with each other in such a way to ensure {\em color
neutrality}\cite{SAT,AM02,GAUSS} over a distance $\Delta x_\perp
\sim 1/Q_s$. Accordingly, the soft gluons with $k_\perp\simle Q_s$
are {\em coherently} emitted from a quasi--neutral gluon
distribution, and then the emission rate ${\partial n}/{\partial Y}$
saturates at a constant value of $\order{1}$. Thus, in the regime
that we call `saturation', the gluon occupation factor keeps
growing, but only {\em linearly} in $Y$ (i.e., as a logarithm of the
energy)\cite{AM99,SAT}. Schematically:

 \beq\label{dndy}
 \frac{\partial n}{\partial Y}\,=\,\chi(n)\,\approx
 \begin{cases}
        \displaystyle{\alpha_s n\,} &
        \text{ if\,  $n\ll {1}/{\alpha_s} \ \ \Longrightarrow \
         n\,\sim\, {\rm e}^{\alpha_s Y}\,, $}
        \\*[0.35cm]
        \displaystyle{\ 1} &  % \frac{1}{\alpha_s}} &
        \text{ if\,  $n\simge {1}/\alpha_s \ \ \ $}
        \displaystyle{
        \Longrightarrow \
        n\,\sim\, Y-Y_s\,,}
        % \frac{1}{\alpha_s}(Y-Y_s)}}
    \end{cases}
  \eeq
where $\chi(n)$ is a non--linear function with the limiting
behaviours displayed above and $Y_s(k_\perp)$ is the saturation
rapidity at which $n(Y_s,k_\perp)\sim 1/\alpha_s$.

The condition $Y
> Y_s(k_\perp)$ for given $k_\perp$ is tantamount to $k_\perp< Q_s(Y)$
for given $Y$, and the occupation number at saturation can be
rewritten as % (cf. Eq.~(\ref{Qsat}))
 \beq n(Y,k_\perp)\,\sim\,Y- Y_s(k_\perp) \,\sim\, \frac{1}{\alpha_s}
 \,\ln\frac{Q_s^2(Y)}{k_\perp^2}\qquad{\rm for}\qquad
 k_\perp< Q_s(Y)\,.
 \eeq
This shows that, due to saturation, the gluon spectrum at low
$k_\perp$ rises only {\em logarithmically} with $1/k_\perp^2$,
instead of the power--like divergence predicted by standard
perturbation theory. (E.g., the bremsstrahlung spectrum (\ref{brem})
would diverge like $1/k_\perp^2$ if extrapolated towards $k_\perp\to
0$.) We see that $Q_s(Y)$ effectively acts as an `infrared cutoff'
in the calculation of the physical observables. This cutoff rises
with the energy, cf. Eq.~(\ref{Qsat}), and also with the atomic
number $A$ in the case where the proton is replaced by a large
nucleus\cite{EdiCGC}: $Q_s^2(Y,A)\sim\rme^{\lambda Y}A^{1/3}$. Hence
for sufficiently high energy and/or large values of $A$, $Q_s^2$
becomes much larger than $\Lambda_{\rm QCD}^2$ and then the
weak--coupling description of the gluon distribution becomes indeed
justified.

Eq.~(\ref{dndy}) is not yet the JIMWLK equation, but only a mean
field approximation to it: In reality, one cannot write down a
closed equation for the 2--point function $n(Y)=\langle
\bm{E}_a\cdot\bm{E}_a\rangle_Y$, rather one has an {\em infinite
hierarchy} for the $n$--point correlations $\langle A(1)A(2)\cdots
A(n)\rangle_Y$ of the gluon fields. In the CGC formalism, these
correlations are encoded into the {\em weight function} $W_Y[A]$ ---
a functional probability density for the field configurations:
 \beq\label{W}
  \langle A(1)\, A(2)\, \cdots\,A(n) \rangle_Y
    \,=\,\int {\rm D}[A]\ W_Y[A]\ A(1)\, A(2)\, \cdots\, A(n)\,.
  \eeq
The average in Eq.~(\ref{W}) is similar to the `average over
disorder' that is usually performed in the study of amorphous
materials, like glasses: the various target configurations scatter
independently with the incoming projectile (indeed, their internal
dynamics is `frozen' over the characteristic time scale for
scattering, by Lorentz time dilation), and the physical scattering
amplitude is finally obtained by summing the contributions from all
such configurations, with weight function $W_Y[A]$. This explains
the concept of `glass' in the `Color Glass Condensate'. The `color'
refers, of course, to the gluon color charge. Finally, the
`condensate' stays for the coherent state made by the gluons at
saturation: this state has a large occupation number $n\sim
1/\alpha_s\gg 1$, as typical for a Bose condensate.

The JIMWLK equation\cite{JKLW,CGC,W} is a {\em functional}
differential equation describing the evolution of $W_Y[A]$ with $Y$.
Via Eq.~(\ref{W}), this functional equation generates an infinite
hierarchy of ordinary evolution equations for the $n$--point
functions of $A$, as anticipated.

\section{DIS off the CGC: Unitarity \& Geometric scaling}
\label{subsec:GEOM}

Let us now discuss the consequences of this non--linear evolution
for the dipole scattering, and thus for DIS. The first observation
is that, when the energy is so high that saturation effects become
important on the dipole resolution scale (this requires $r\simge
1/Q_s(Y)$, cf. Eq.~(\ref{T1SCATT})), then {\em multiple scattering}
becomes important as well: e.g., the double--scattering $T^{(2)}\sim
(\alpha_s n)^2$ is of $\order{1}$, so like the single--scattering
$T^{(1)}\sim \alpha_s n$. Thus, the behaviour of the scattering
amplitude in the vicinity of the unitarity limit is {\em the
combined effect of BFKL growth, gluon saturation and multiple
scattering}.

Within the CGC formalism, multiple scattering is easily included in
the {\em eikonal approximation}, thus yielding
 \beq\label{avT}  \langle T(r,b)\rangle_Y\,=\,\int {\rm D}[A] \
 \,W_Y[A]\,\,\,T(r,b)[A]\,,\eeq
where $T[A]$ is the amplitude corresponding to a given configuration
of classical fields $A$, and is non--linear in the latter to all
orders. By taking a derivative w.r.t. $Y$ and using the JIMWLK
equation for $\partial W_Y/\partial Y$, one can deduce an evolution
equation for the (average) dipole amplitude, with the following
schematic structure (we ignore the transverse coordinates) :
  \beq\label{B1}
  \del_Y\langle T\rangle\,=\,\alpha_s\langle T\rangle \,-\,
 \alpha_s\langle T^2\rangle\,.\eeq
Note that this is not a closed equation --- the amplitude $\langle
T\rangle$ for one dipole is related to the amplitude $\langle
T^2\rangle$ for two dipoles --- but only the first equation in an
infinite hierarchy, originally obtained by Balitsky\cite{B}. A
closed equation can be obtained if one assumes factorization:
$\langle T^2\rangle\approx \langle T\rangle \langle T\rangle$. This
{\em mean field approximation} yields the Balitsky--Kovchegov (BK)
equation\cite{K}, which applies when the target is sufficiently
dense to start with (so like a large nucleus) and up to not too high
energies (cf. Sec. \ref{sec:PLOOP}).

Due to its simplicity, the BK equation has played an important role
as a laboratory to study the effects of saturation and multiple
scattering. As already manifest on its schematic form in
Eq.~(\ref{B1}), this equation has the fixed point $\langle
T\rangle=1$ at high energy, i.e., it is consistent with unitarity
and, moreover, it predicts that the black disk limit is eventually
saturated: $\langle T(r)\rangle_Y = 1$ when $r\simge 1/Q_s(Y)$.
Hence, one can use this equation to determine the energy--dependence
of the saturation momentum, that is, the slope $\lambda$ of the
saturation line $\ln Q^2_s(Y)=\lambda Y$ (cf. Fig.
\ref{HERA-gluon}). Remarkably, the growth of $\langle T\rangle_Y$
with $Y$ before reaching the saturation is entirely determined by
the linearized version of the BK equation, i.e., the BFKL equation.
This is important since, unlike the BK equation, the BFKL equation
is presently known to NLO accuracy\cite{NLBFKL}. By using the latter
(within the collinearly improved NLO--BFKL scheme of
Refs.\cite{Salam99}), Triantafyllopoulos has computed\cite{DT02} the
{\em saturation exponent} $\lambda$ to NLO accuracy and thus found a
value $\lambda\simeq 0.3$, which is roughly one third of the
corresponding LO estimate\cite{GLR}.

\begin{figure}
\begin{center}\hspace*{-.1cm}
       \mbox{{\epsfig{file=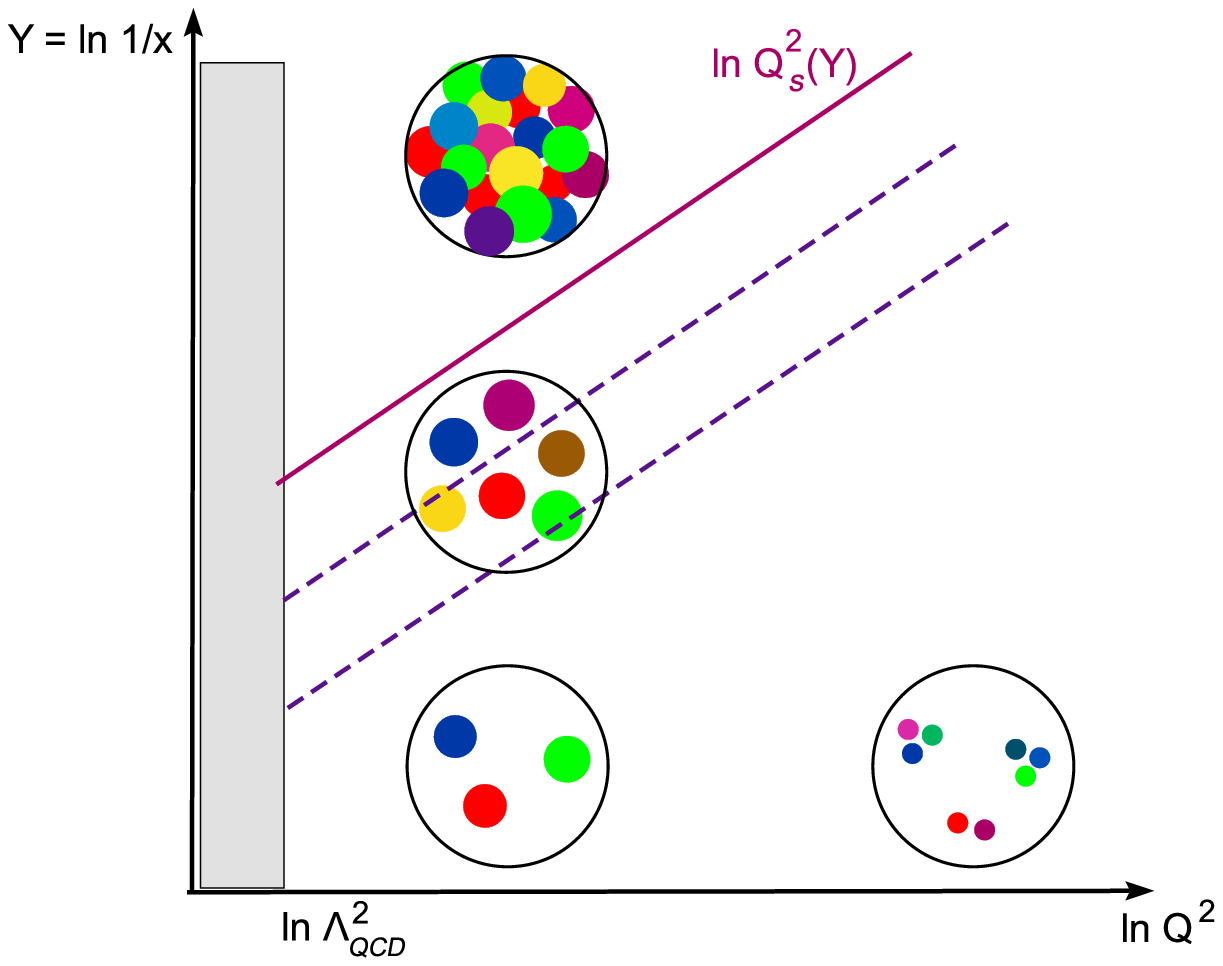,
        width=0.6\textwidth}}%\hspace*{-1.cm}
             {\epsfig{figure=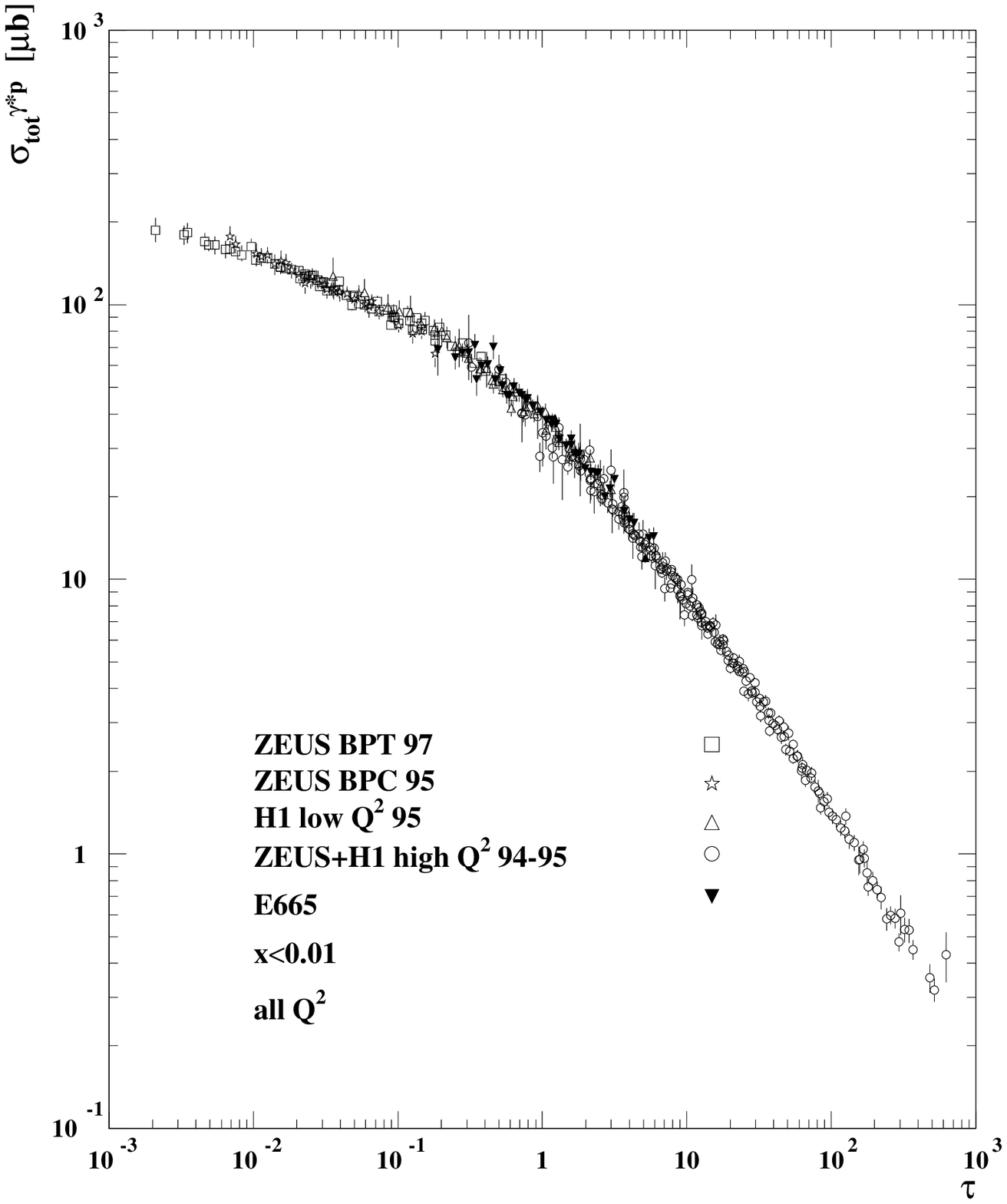,
        width=0.47\textwidth}}}    %\vspace*{-.4cm}
\caption{\sl Left: Line of constant gluon occupancy at high $Q^2$.
Right: Geometric scaling in the HERA data for $\sigma_{\gamma^*p}$
at $x\le 0.01$; $\tau$ is the scaling variable, $\tau\equiv
Q^2/Q^2_s(Y)$.} \label{HERA-scaling}
\end{center}\vspace{-.5cm}
\end{figure}

Another crucial consequence of the non--linear evolution towards
saturation --- at least, at the level of the BK equation --- is the
property known as {\em geometric scaling\,}\cite{geometric,SCALING}
: Physics should be invariant along trajectories which run parallel
to the saturation line because these are lines of {\em constant
gluon occupancy} (see the l.h.s. of Fig. \ref{HERA-scaling}). In
mathematical terms, this means that, up to relatively large momenta
$Q^2 \gg Q_s^2(Y)$, the observables should depend only upon the
difference $\ln Q^2 - \ln Q^2_s(Y)=\ln [Q^2/Q^2_s(Y)]$ from the
saturation line, i.e., they should {\em scale} upon the ratio
${\tau}\equiv Q^2/Q^2_s(Y)$ rather than separately depend upon $Q^2$
and $Y$. In particular, the dipole scattering amplitude obeys
$\langle T(r)\rangle_Y\approx \mathcal{T}(r^2Q^2_s(Y))$, which via
the factorization formula (\ref{dipolefact}) implies a similar
scaling for the DIS cross--section (in the limit where the quark
masses are negligible) :
$\sigma_{\gamma^*p}(Y,Q^2)\approx\sigma({\tau})$.

Remarkably, such a scaling has been identified in the HERA data, by
Sta\'sto, Golec-Biernat and Kwieci\'nski\cite{geometric} (see the
r.h.s. plot in Fig. \ref{HERA-scaling}, which is taken from
Ref.\cite{geometric}), before its theoretical explanation has
emerged\cite{SCALING,MT02,MP03} from studies of the BK equation.
More recently, with the advent of more precise data for DIS
diffraction at HERA, geometric scaling has been noticed in these
data too\cite{MS06}.

The outstanding feature of this scaling is the fact that this is a
consequence of saturation which manifests itself up to relatively
large transverse momenta, well above the saturation
scale\cite{SCALING}. This is consistent with the HERA data, which
show approximate scaling for all the experimental points at $x <
0.01$, including values of $Q^2$ as high as 400 GeV$^2$. (For
comparison, the saturation scale estimated from these data is
$Q_s\simeq 1$ GeV for $x\sim 10^{-4}$.) It is also interesting to
notice that the value for the saturation exponent coming out from
such scaling fits to HERA is in agreement with its theoretical
estimate\cite{DT02} $\lambda\simeq 0.3$. Moreover, the {\em
violations} of geometric scaling observed in the HERA data appear to
be consistent\cite{IIM03} with theoretical expectations from the
BFKL dynamics\cite{SCALING,MT02}.

The study of the BK equation has led to another surprise: Munier and
Peschanski recognized\cite{MP03} that this equation is in the same
universality class as the FKPP equation which describes the mean
field limit (corresponding to very large occupation numbers) of the
classical stochastic process known as {\em
reaction--diffusion}\footnote{The reaction--diffusion process $A
\rightleftharpoons AA$ can be briefly described as
follows\cite{Saar} : `molecules' of type $A$ which are located at
the sites of an infinite, one--dimensional, lattice can locally
split ($A\to AA$) or merge ($AA\to A$) with each other; also, a
molecule can diffuse to the adjacent sites. In the analogy with QCD,
the `molecules' correspond to gluons, the one--dimensional `spacial'
axis is the logarithm of the gluon transverse momentum, the
`particle splitting' corresponds to the BFKL evolution, and the
`particle merging' to the non--linear effects responsible for gluon
saturation. Finally, the `diffusion' corresponds to the
non--locality of the BFKL kernel and the various gluon vertices in
transverse space.}. This observation shed a new light on the physics
of geometric scaling and, moreover, it helped clarifying the
limitations of the mean field approximations and the essential role
of {\em fluctuations}. In turn, this lead to substantial theoretical
progress over the last two years that I shall briefly discuss in the
next section.

\section{Gluon number fluctuations and pomeron loops}
\label{sec:PLOOP}

Although the experimental results at HERA, and also at RHIC (see,
e.g., the recent reviews in Ref.\cite{CGCRHIC}), appear to be
consistent with theoretical expectations based on the BK or JIMWLK
equations\footnote{The BK equation represents the large--$N_c$ limit
of the Balitsky--JIMWLK hierarchy.}, the latter are nevertheless
incomplete (even to lowest order in $\alpha_s$) and thus cannot
describe the actual dynamics in QCD at very high energies. Indeed,
as recently recognized in Refs.\cite{IM032,MS04,IMM04,IT04}, these
equations miss the effects of {\em gluon--number fluctuations} in
the dilute regime, which however have a strong influence on the
evolution with increasing energy, in particular, on the dynamics
towards saturation. Such a strong sensitivity to fluctuations may
look at a first sight surprising --- the high--energy regime is
characterized by high gluon occupancy, and therefore should be less
affected by fluctuations ---, but in fact this was already noticed
in early studies of unitarization in the context of the dipole
picture\cite{AM94} and, more recently, it has been rediscovered
within the context of the non--linear QCD evolution in the vicinity
of the saturation line\cite{IM032,MS04}. This is also in
agreement\cite{IMM04} with known properties of the
reaction--diffusion process, as originally discovered in the context
of statistical physics\cite{Saar}. There are several ways to
understand this sensitivity to fluctuations:

\texttt{(i)} First, one may recall from Fig. \ref{EvolREC} that
non--linear phenomena like gluon saturation and multiple scattering
involve the simultaneous exchange of several gluons in the
$t$--channel, and thus they probe {\em correlations} in the gluon
distribution. At high energy, the most important such correlations
are those generated via {\em gluon splitting} in the dilute regime:
the `child' gluons produced after a splitting are correlated with
each other because they `remember' about their common parent. These
correlations manifest themselves in the difference $\langle
nn\rangle - \langle n\rangle \langle n\rangle$ between the average
pair density $\langle nn\rangle$ and its mean--field piece $\langle
n\rangle \langle n\rangle$; thus, they describe {\em gluon--number
fluctuations}\cite{AM94,IM032,IT04}. Alternatively, these
correlations are responsible for the difference $\langle T^2\rangle
-\langle T\rangle \langle T\rangle$ (cf. Eq.~(\ref{B1})) and hence
for {\em violations} of the factorization assumption underlying the
BK equation\cite{IM032,MS04}.

\texttt{(ii)} Second, one may notice that the driving force behind
the high--energy evolution is the BFKL growth in the {\em dilute
tail} of the gluon distribution at large transverse momenta
$k_\perp\gg Q_s(Y)$. In that tail, the gluon occupation numbers are
still low ($n\ll 1/\alpha_s$), so the fluctuations are relatively
important, as demonstrated by intensive studies in the context of
statistical physics (see Ref.\cite{Saar} for a recent review and
more references), which in turn have inspired similar studies within
QCD\cite{IMM04,GS05,EGBM05}.

Whereas the failure of the BK equation to accommodate these
fluctuations was a priori clear, it somehow came as a
surprise\cite{IT04} that a similar failure holds also for the more
general, Balitsky--JIMWLK, equations. The original confusion on this
point came from the fact that these equations do generate
correlations, as obvious from the fact that they correspond to
non--trivial hierarchies. However, it turns out that the respective
correlations represent {\em color} fluctuations alone, and thus
disappear in the limit where the number of colors is large, $N_c\gg
1$, unlike the fluctuations in the particle number.

After this failure has been recognized\cite{IT04}, new equations
have been proposed\cite{IT04,MSW05,IT05}, which encompass both
saturation and fluctuations in the limit where $N_c$ is large. These
equations have been interpreted\cite{LL05,BIIT05,IST05} as an
effective theory for BFKL `pomerons', in which the pomerons are
allowed to dissociate and recombine with each other, like the
molecules in the reaction--diffusion problem. Thus the perturbative
solution to these equations involves {\em pomeron loops}. But the
complexity of these `pomeron loop' equations has so far hindered any
systematic approach towards their solutions, including via numerical
methods. The only properties of these solutions to be presently
known come essentially from the correspondence with statistical
physics\cite{MP03,MS04,IMM04,IT04,GS05,EGBM05}, which is however
limited to asymptotically high energies and very small values of the
coupling constant.  The structure of these equations together with
the known results about their solutions are discussed in more detail
in other talks at this conference\cite{DISCGC}.

\begin{figure}[t]%\hspace*{-1.cm}
    \centerline{\epsfxsize=10.cm\epsfbox{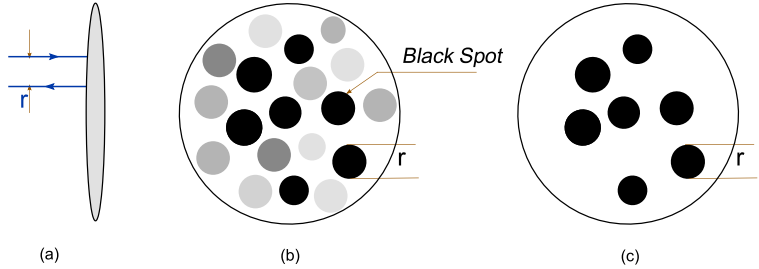}}
    \caption{\sl Dipole--hadron scattering in the
    fluctuation--dominated regime at very high energy.
    (a) a view along the collision axis; (b)
    a transverse view of the hadron, as `seen' by a small dipole
    impinging at different impact parameters; (c) the simplified,
     black\&{white}, picture of the hadron which is relevant for
     the average dipole amplitude.
    \label{HotSpot}}%\bigskip
    \end{figure}

Fig. \ref{HotSpot} illustrates the most striking consequences of the
evolution with pomeron loops, as probed in DIS at very high energy.
The small blobs which are grey or black represent the regions of the
target disk which are explored by the dipole with size $r$ at
various impact parameters. The nuance of grey is representative of
the intensity of the interaction, and thus of the local gluon
density in the target (as viewed on the resolution scale $Q^2=1/r^2$
of the incoming dipole): a light grey spot denotes weak scattering
($T(r,b)\ll 1$), and hence a region with low gluon density, a white
region means almost no gluons at all, and a black spot represents,
of course, a region where the gluon density is so high that the
black disk limit is reached: $T(r,b)\approx 1$. Thus, this picture
suggests that, when probed on a fixed resolution scale $Q^2$, a
hadron at very high energy may look extremely inhomogeneous. This
lack of homogeneity has nothing to do with the initial conditions at
low energy, rather it is the result of gluon--number fluctuations in
the high--energy evolution. What is most remarkable about this
picture is that, for sufficiently high energy, the {\em average}
amplitude $\lan T(r)\ran_Y$ (and thus the DIS cross--section) is
completely dominated by the black spots up to very large values of
$Q^2$, well above the {\em average} saturation momentum $\lan
Q_s^2\ran_Y$. That is, the average scattering may be weak, $\lan
T(r)\ran_Y\ll 1$, meaning that the target looks dilute {\em on the
average}, yet this average is in fact controlled by {\em rare
fluctuations} with unusually large density, for which $T\sim 1$.
From the perspective of the incoming dipole, the hadron disk looks
either black ($T\simeq 1$) or white ($T\simeq 0$), as illustrated in
Fig. \ref{HotSpot}.c.

This physical picture has interesting consequences for the
phenomenology. For instance, for DIS it predicts that, at
sufficiently high energy, geometric scaling should be washed out by
fluctuations\cite{MS04} and replaced by a new type of
scaling\cite{IMM04,IT04}, known as {\em diffusive
scaling}\cite{HIMST06}\,: instead of being a function of the ratio
$Q^2/Q^2_s(Y)$, the DIS cross--section at high--energy should rather
scale as a function of $Z\equiv \ln [Q^2/Q^2_s(Y)]/\sqrt{DY}$. (A
similar scaling holds for the diffractive
cross--section\cite{HIMST06} in DIS and also for the cross--section
for gluon production in proton--proton scattering at forward
rapidity\cite{GLUON}.) Here, $D$ is a `diffusion coefficient' which
measures the dispersion in the gluon distribution due to
fluctuations. Its value is in principle determined by the pomeron
loop equations, but it is presently unknown, because of the lack of
a explicit solutions to the latter. Knowing the actual value of this
parameter would be extremely important, since this would tell us at
which energy we should expect diffusive scaling. (We
expect\cite{HIMST06} geometric scaling for $DY\ll 1$ and diffusive
scaling for $DY\simge 1$.)

\begin{figure}[t]
    \centerline{\epsfxsize=8.5cm\epsfbox{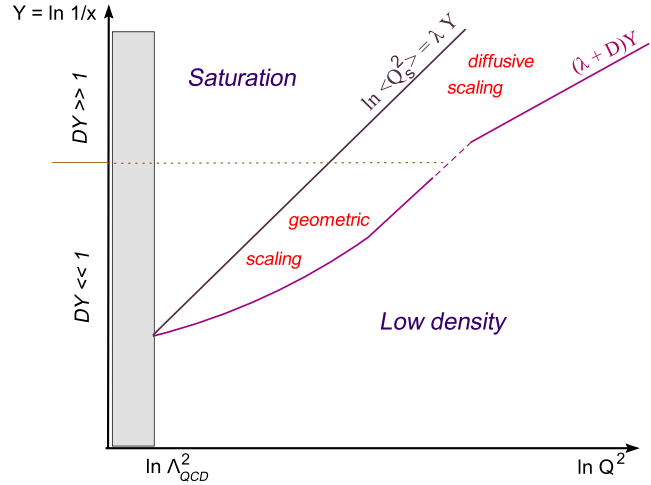}}
    \caption{\sl A `phase--diagram' for the high--energy evolution
    with pomeron loops. Shown are the average saturation line and
    the scaling regions at $Q^2> \lan Q^2_s\ran_Y$.
    \label{phase}}%\vspace{-.2cm}
    \end{figure}

The previous discussion is summarized by the `phase--diagram' in
Fig. \ref{phase}, which exhibits more structure than the original
one in Fig. \ref{HERA-gluon}: In addition to the saturation line (to
be now understood as the {\em average} saturation line $\ln \,\lan
Q^2_s\ran_Y\!=\lambda Y$ in the presence of fluctuations), this
diagram also shows the kinematical domains for geometric scaling (at
intermediate energies: $DY\ll 1$) and, respectively, diffusive
scaling (at very high energy: $DY\simge 1$), which are seen to
extend up to relatively large $Q^2\gg \lan Q^2_s\ran_Y$.  This is
important as it shows that, for sufficiently high energies, the
physics of saturation should manifest itself as the breakdown of the
standard approximations at high $Q^2$ (like the leading--twist
approximation or the collinear factorization) up to values of $Q^2$
which are so large that the {\em average} scattering amplitudes are
truly small, far below the `black disk limit'.

The precise locations of the borderlines between these various
regimes are not fully under control, because of the theoretical
uncertainties on $\lambda$ and $D$. However, the experimental
results at HERA and RHIC suggest that these experiments may have
already probed the intermediate energy range characterized by
geometric scaling (although in a kinematical domain which is only
marginally perturbative). The experimental situation at the LHC will
be even more favorable in that respect. The energies to be available
there will be so high that the physics of saturation and the CGC
could be explored within a wide kinematical range, including
relatively large values of $Q^2$ for which the perturbation theory
is fully reliable. In particular, there is the interesting
possibility that the results at LHC will capture the transition from
geometric to diffusive scaling (e.g., by varying the rapidity of the
particles produced in $pp$ or $pA$ collisions\cite{GLUON}), and thus
unveil the ultimate regime of QCD at ultrahigh energies.

%\bibliographystyle{unsrt}
%\bibliography{myrefs}

\end{document}